\documentclass[12pt]{article}
\usepackage{graphicx}

\newcommand{\beq}{\begin{equation}}
\newcommand{\eeq}{\end{equation}}
\newcommand{\bea}{\begin{eqnarray}}
\newcommand{\eea}{\end{eqnarray}}

\begin{document}
\begin{center}
\noindent {\Large Multifractal Fingerprints in the Visual Arts}
\vskip .3cm
\noindent {J.\ R.\ Mureika \\
{\footnotesize Email: jmureika@jsd.claremont.edu \\
{\it W.\ M.\ Keck Science Center, The Claremont Colleges}
\\ {\it 925 N. Mills Avenue, Claremont, California~~91711-5916}}}
\vskip .3cm
\noindent {G.\ C.\ Cupchik \\
{\footnotesize Email: cupchik@scar.utoronto.ca \\
{\it Division of Life Sciences, University of Toronto at Scarborough}
\\ {\it 1265 Military Trail, Scarborough, ON~~Canada~M2C 1A4}}}
\vskip .3cm
\noindent {C.\ C.\ Dyer \\
{\footnotesize Email: dyer@astro.utoronto.ca \\
{\it Department of Astronomy and Astrophysics, University of Toronto} 
\\ {\it 60 St.\ George Street, Toronto, ON~~Canada~M5S 3H8}}}
\end{center}
\vskip .5cm
\noindent Abstract\\ {\footnotesize The similarity in fractal dimensions
of paint ``blobs'' in samples of gestural expressionist art implies that
these pigment structures are statistically indistinguishable from one
another. This result suggests that such dimensions cannot be used as a
``fingerprint'' for identifying the work of a single artist.  To overcome
this limitation, the multifractal spectrum is adopted as an alternative
tool for artwork analysis.  For the pigment blobs, it is demonstrated that
this spectrum can be used to isolate a construction paradigm or art style.  
Additionally, the fractal dimensions of edge structures created by
luminance gradients on the canvas are analyzed, yielding a potential
method for visual discrimination of fractally-similar paintings. }

\pagebreak

The notion that Nature can be described by fractal geometry was first
suggested by Benoit Mandelbrot \cite{mandel}.  A fractal is a recursive,
self-similar structure whose constituent parts in some way resemble the
whole.  Mathematically, these are defined as

\begin{equation}
N(d) \propto d^{D_F}~,
\label{eqn1}
\end{equation}

Here, $N(d)$ is a measure of the number of ``objects'' which comprise the
set at a viewing scale $d$,
and $D_F$ is the fractal dimension, which can be interpreted as a
measure of the irregularity of the structure.  Simple, smooth shapes such
as dots, circles, or spheres have fractal dimensions which coincide with
their Euclidean dimension.  Conversely, a fractal is defined by a
non-integer dimension which acts as a measure of the object's roughness.  
For example, a line has dimension 1, but if it becomes very jagged at many
scale levels, its dimension rises fractionally above this value - it has
become a fractal.  If the line becomes so jagged and rough that it
effectively fills an area, then it has a fractal dimension approaching 2.

A recent surge of interest in fractal geometric ``fingerprinting'' of
natural phenomena has included the study of artistic style.  The style in
question is gestural expressionism, a mid-twentieth century technique in
which the artist's hand movements are guided by a philosophy of psychic
automatism and the resulting images are seemingly disordered and chaotic.
Researchers have proposed that a characteristic fractal dimension may be
associated with the work of Jackson Pollock, identifying the physical
distribution of pigment patterns as the associated 
fractal \cite{taylor1,taylor2,taylor3,taylor4}. If a
distinctive fractal dimension could be uncovered for every artist, this
would pave the way for a novel form of artwork authentication. But is this
technique sufficient to distinguish artists within the same gestural
expressionist group, in this case between Jackson Pollock and the Quebec
Automatistes, including artists such as Marcel Barbeau and Jean-Paul
Riopelle?

A comparison was made between two groups of 8 paintings by Pollock and Les
Automatistes.  The images were digitized as 24-bit color files of sides
ranging between 1000-2500 pixels, and pigment patterns were filtered out
according to a specified target color in RGB space (see Figures 1 and 2).  
A variance in the values of the R, G, and B channels (each between 0-255
for 24-bit color) up to a specified distance from the target was
allowed to account for any small fluctuations in the pigment shade.  The
fractal dimensions of the resulting patterns were calculated by the
standard box counting method covering roughly 3 orders of magnitude of
scale (1000 pixels to 4 pixels per side), roughly several meters to a few
millimeters in terms of the actual canvas dimensions.  As the patterns are
a result of random monochromatic pigment deposits we hereafter refer to
them as ``blobs'', an etymology based on the ``elongated blobs'' of Julesz as
distinguishable perceptual objects \cite{julesz}.

A one-way analysis of variance comparing the 8 Pollock (mean $D_F = 1.79$)
with the 8 Automatistes ($D_F = 1.73$) paintings indicated that the $D_F$
indices were not significantly different, $F(1,14)=1.18~p < .30$. This
suggests that the fractal dimension of drip paintings is not unique to any
one artist and cannot be used for any such type of authentication scheme.  
It should be noted that a more recent study \cite{taylor3} has found that a fractal
box counting analysis can differentiate between five Pollock and five
non-Pollock images.  These results can be considered to be consistent with
those reported in the paper, since the non-Pollock images could be painted
in such a way as to be ``non-gestural''.  Future analysis can shed more
light on this finding.

Since a single fractal dimension rarely represents the true structure of
natural objects, the multifractal spectrum of an image may provide a more
rigorous way to classify the style or construction paradigm of paintings
associated with a particular group such as gestural expressionists.  A
multifractal is a set whose form is a weave of overlapping self-similar
configurations.  These geometric formulations have been shown to describe
the physical organization of a myriad of natural phenomena, ranging from
tree root growth to large-scale galaxy clustering \cite{vicsek,piet}.  Unlike simple
fractals, multifractals are characterized by an infinite set of dimensions
$\{D_q\}=\{D_0,D_1,D_2,\ldots\}$, calculated in a similar manner to 
$D_F$, which determine the scaling
structure as a function of the local pattern density.  The subscript $q$ is
generally an integer, where $q = 0$ represents the classic fractal dimension
($D_F = D_0$).  The regions of densest clustering, represented 
by extremely large values of $q$ (or $q\rightarrow \infty$) scale
according to the dimension $D_\infty \leq D_0$.  These two statistics, 
and all those in
between, give a much deeper insight into the physical organization of the
object in question, and in fact can be used as a method of identifying the
associated formation mechanism (as was discussed in \cite{mephd}).  In the case of
a regular fractal, all multifractal dimensions $\{D_q\}$ are equal to $D_F$.

The multifractal spectra ${D_q}$ were determined for the 8 Pollock (mean
$D_0=1.60$) and 8 Automatistes (mean $D_0 =1.58$) paintings, and a one-way
analysis of variance reveals that these do not present a clearer means of
differentiation than the base dimension, $F(1,14) = 0.06,~p < .80$. The
gestural expressionist paintings by Pollock and the Automatistes were then
compared with 6 paintings chosen from a different style, Artonomy or
Systematic Art, created by Tsion Avital \cite{avital1}.  
This alternate technique
involves creating paintings which are grouped in series according to
strict rules of transformation and are ``meaningless'' when individually
taken out of context. However, the individual paintings simply serve as a
control in this study. A sample of Avital's Artonomy is shown in Figure 3.

For 6 of these systematic art images, the analysis derived a mean $D_F=1.60$
and $D_\infty=1.58$, suggesting instead a monofractal structure ($D_F = D_\infty$).  
This should be compared with the multifractal ``depth'' of the Pollock and
Automatistes works which show mean differences $D_F - D_\infty =$ 0.19 and 0.15,
respectively.  This indicates that the set of dimensions $\{D_q\}$ of paint
blobs can be used only to differentiate between ``classes'' of painting but
not conclusively between different artists within the same class.  The
multifractal spectrum is interpreted here as the signature of an artistic
style \cite{mephd}.

How then can one distinguish between artists within particular stylistic
groups?  It was noted that humans have a preference for fractals
dimensions of about 1.8 \cite{taylor1}, suggesting that the gestural expressionists
catered their craft to this special dimension. However, according to
Berlyne \cite{berlyne1} test subjects were found to have a visual propensity
for images which are less complex, or contain more symmetric and
heterogeneous information.  This fact was more recently confirmed
independently by Taylor \cite{taylor4}, who report that human visual
preference is tuned to $D_F \sim 1.3$.  The images deemed ``pleasing'' 
in reference \cite{berlyne1} consist of regularly overlapping Euclidean shapes, which would
suggest a fractal dimension closer to (but greater than) 1.  This poses
the very interesting question of why these artists gear their paintings to
such high values if they are not deemed ``perceptually favorable''?

Moreover, if there is no appreciable difference between the base fractal
statistics for the pigment distributions, as in the case of the images in
Figures 1 and 2, what is it about the paintings that can impart different
visual sensations?  In their seminal work, Hubel and Weisel (see e.g. \cite{hubelweisel})
have established the principle that the brain is naturally disposed to
analyze visual structures in terms of edges.  A study of these edges on
the canvases should thus reveal new information about the perceptual
nature of the artworks.

The standard RGB primary color decomposition can be seen as a reflection
of the eye's sensitivity to specific wavelengths of light via the L, M,
and S cone cells.  Following the notion of edge detectors in the brain, it
makes sense to approach the problem in a different color space
representation, namely YIQ.  This offers an alternative method to
decompose chromaticity information in terms of luminance (Y), hue (I), and
saturation (Q) instead of red, green and blue primaries (see e.g. \cite{cgraphics} for
further details on color spaces).

The paintings by Jackson Pollock, Les Automatistes, and Tsion Avital were
compared in terms of luminosity gradients, which were obtained by applying
a Sobel filter to the luminance channel (Y) whose values again range
between 0 (black, no gradient) to 255 (white, high gradient).  The edge 
structure is defined as the
regions of strongest color contrast and the associated fractal dimensions
$D_F$ for each painting were obtained.  In this case, the $D_F$ of these
patterns showed decided grouping, unlike those of the physical paint
blobs.  A highly significant difference was found for Pollock's edges ($D_F
= 1.84$) as compared to those of Les Automatistes ($D_F = 1.48$), $F(1,14)
=14.52,~p < .002$.  The works of Jackson Pollock thus show highly irregular
edge structures (characterized by $D_F$ close to 2), compared with those of
Les Automatistes, while Artonomy's edges possess very simple Euclidean
organization ($D_F$ roughly 1).  It is therefore the irregularity of 
edges that makes
Jackson Pollock's style unique and is representative of the degree of
``expressionism'' in the painting.

The suggestion that patterns of similar fractal dimension are perceptually
indistinguishable can be related to the work of Julesz \cite{julesz} who argues
that texture discrimination in ``effortless'' or ``immediate'' perception can 
only occur for configurations whose autocorrelation power spectra are
different.  Such correlation statistics can be implicitly linked to the
multifractal spectrum ({\it e.g.} $D_2$ is equivalent to the two-point 
correlation exponent, a structural measure of ``pair-clustering'' between 
points on the image) and thus these conclusions provide a natural extension of
earlier findings.

The juxtaposition of blobs versus edges on the canvas provides two
distinct structures in one painting.  One facet of the image results from
the deposits of raw pigment on the canvas, while another facet has as its
origins the boundary between two adjacent colors.  The similarity between
these definitions and one of Julesz's fundamental classes of topological
perception units (coined textons in \cite{julesz,julesz2}, analogous to what we term
blobs) lends further support to the idea that these are visually
discriminable patterns.  In fact, as the density of edges on the canvas
increases, the edge structures themselves become blobs.  Mathematically,
this is evident in the sense that $D_F$ (or $D_0$) are very close to 2.

A two-way mixed model ANOVA was conducted treating Artist (Pollock/Les
Automatistes) as a between-subjects variable and Structure (blob/edge) as
a within-subjects variable. A significant two-way interaction of Artist
and Structure, $F(1, 14) = 5.51, p < .03$, shows that the blobs and the edges
that they create are not significantly different for Pollock (see Figure
4).  This equivalence of mean fractal dimensions implies a ``symmetry''
between indistinguishable components which form a cohesive whole. However,
for the Automatistes, the density of edges is significantly less than the
density for blobs.  There is thus a breakdown in the structural symmetry
in this case, yielding a perceptual ``conflict'' of two nested but
distinguishable characteristics of the painting.  Thus, for the Pollock
paintings, the viewer transitions effortlessly between blobs and edges,
but not so for Les Automatistes.

The contrast between edges and blobs has figured prominently in art
historical analysis as the ``linear versus painterly dimension'' discussed
in \cite{wolfflin}. While the linear is characteristic of classical art styles which
favor clear edges and structured space, the sketchy baroque and
impressionist styles are more painterly, encouraging viewers to complete
an image. The linear versus painterly dimension has also consistently
emerged as the primary one underlying perceptual discriminations between
pairs of paintings. This applies to paintings selected across a broad
spectrum of traditions \cite{berlyne2,gerry1} as well as those produced by 
Avital's Systematic Art approach \cite{cupchikavital}.

In sum, this new study has shown that Jackson Pollock is unique within
gestural expressionism because of the irregularity or degree of roughness
of edges underlying the structure of his paintings. It is precisely the
disposition of the brain to discriminate edges [8] that makes it so
sensitive to this fractal property in Jackson Pollock. The fact that
fractality is a property of the whole implies that order is discerned in
seeming chaos and this may provide a foundation for the pleasure
experienced by some when viewing his paintings.

\vskip .5cm
\noindent{\bf Acknowledgments}\\
This work is supported by grants from the Natural Sciences and Engineering
Research Council of Canada.  The image {\it Reflections of the Big Dipper
(1947)} by Jackson Pollock (Figure 1(a)) was provided by 
Art Resource, NY, and has been reproduced with the
permission of the Artists Rights Society.  We
graciously thank Tsion Avital for the permission to reproduce his works in
Figure 3.

\pagebreak
%

\noindent {\bf Copyright credits:}\\
{\it Reflections of the Big Dipper (1947)}, Pollock, Jackson (1912-1956) \\
$\copyright$ 2003 The Pollock-Krasner Foundation/Artists Rights Society (ARS),
New York;
Stedelijk Meuseum, Amsterdam, The Netherlands

\begin{figure}[h] \begin{center} \leavevmode
\includegraphics[width=0.5\textwidth]{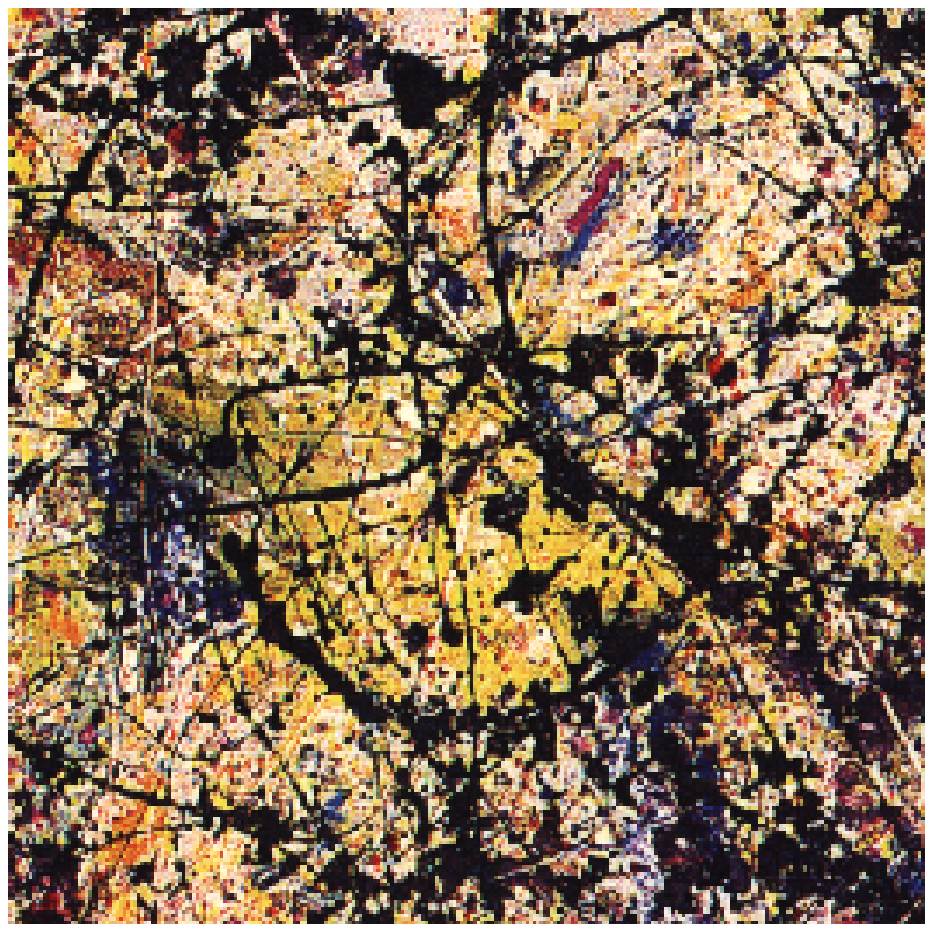}~~~
\includegraphics[width=0.5\textwidth]{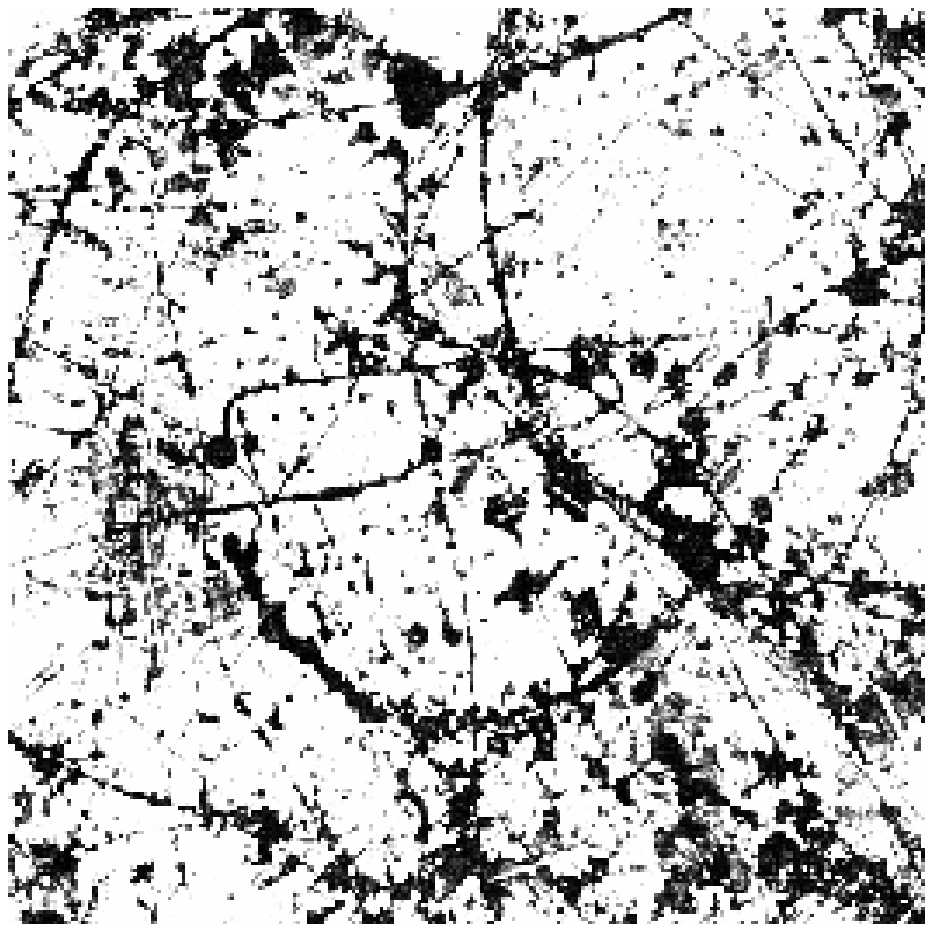} 
\vskip .3cm
\includegraphics[width=0.5\textwidth]{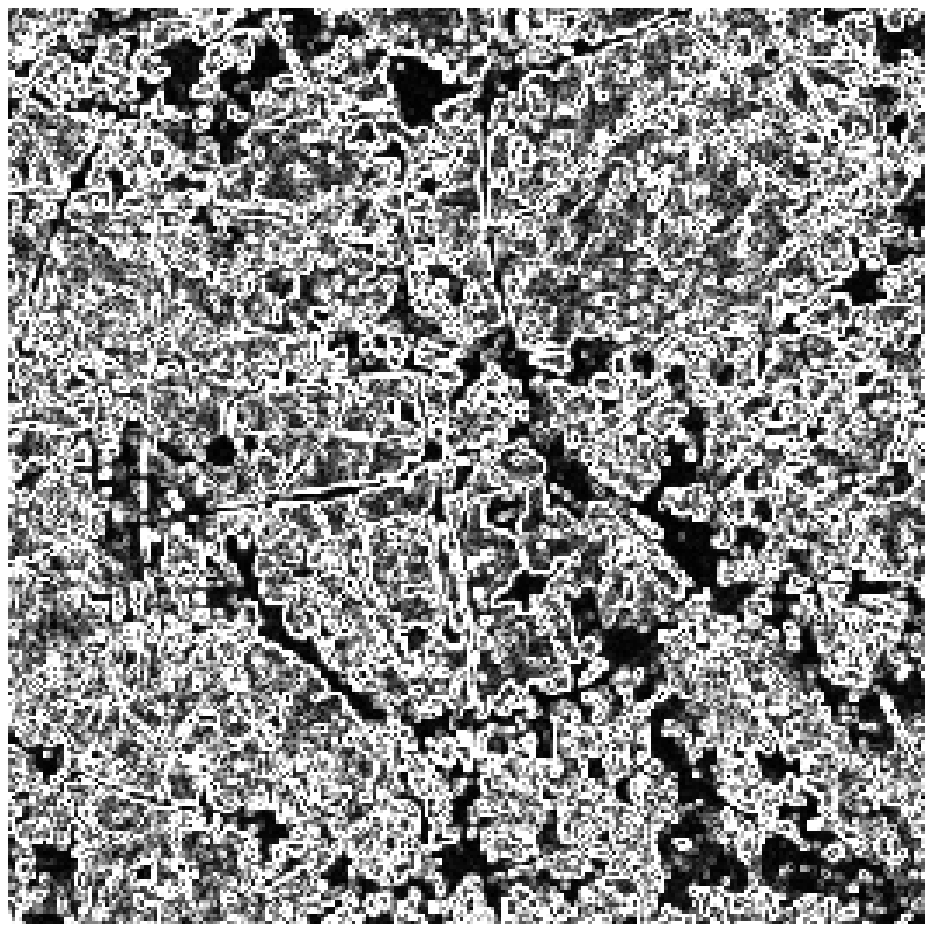}
\end{center} 
\caption{
{\it Reflections of the Big Dipper (1947)}, by
Jackson Pollock.
Image progression shows (A) original painting, (B) blob
structure (black pigment) and (C) luminance edge structure (white regions). 
}
\label{figure1}
\end{figure}

\pagebreak

\begin{figure}[h] \begin{center} \leavevmode
\includegraphics[width=0.5\textwidth]{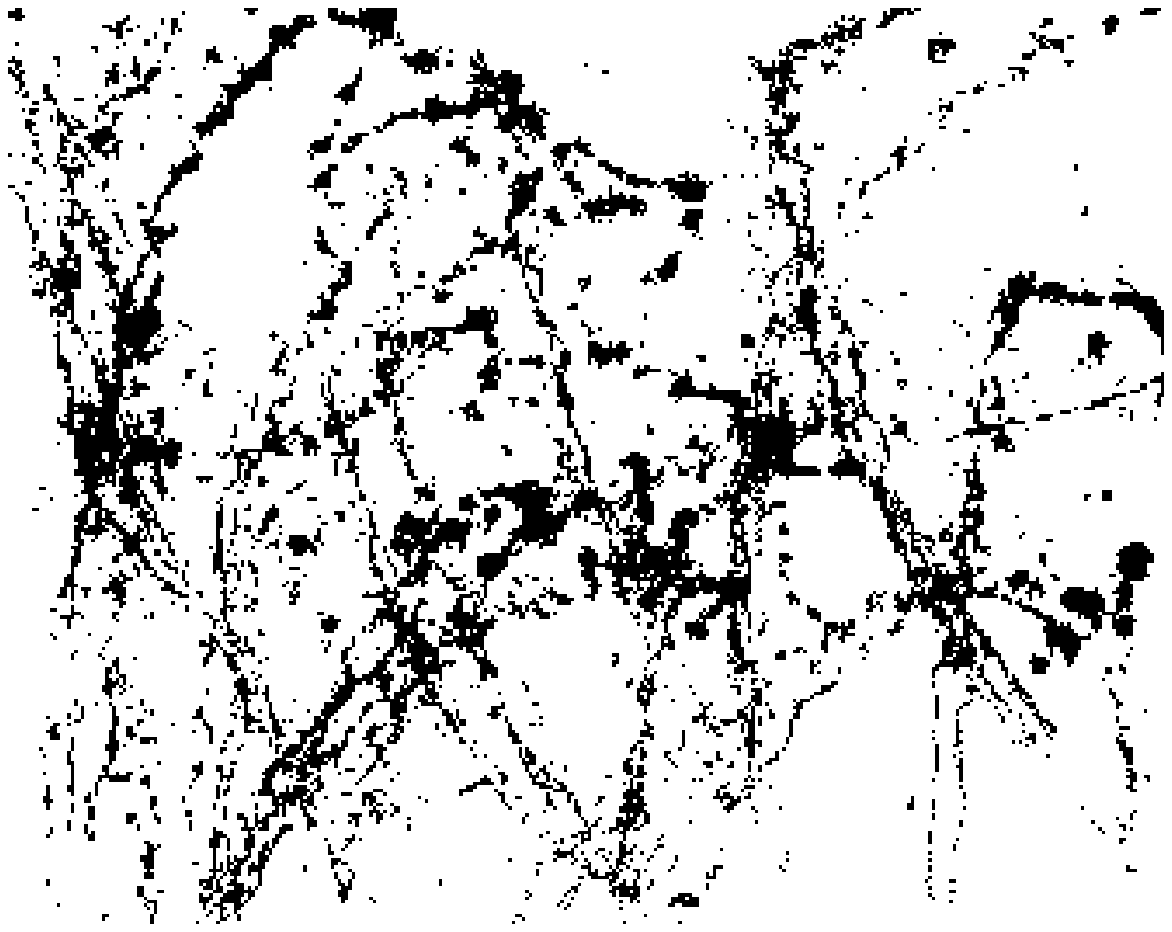}
\vskip .3cm
\includegraphics[width=0.5\textwidth]{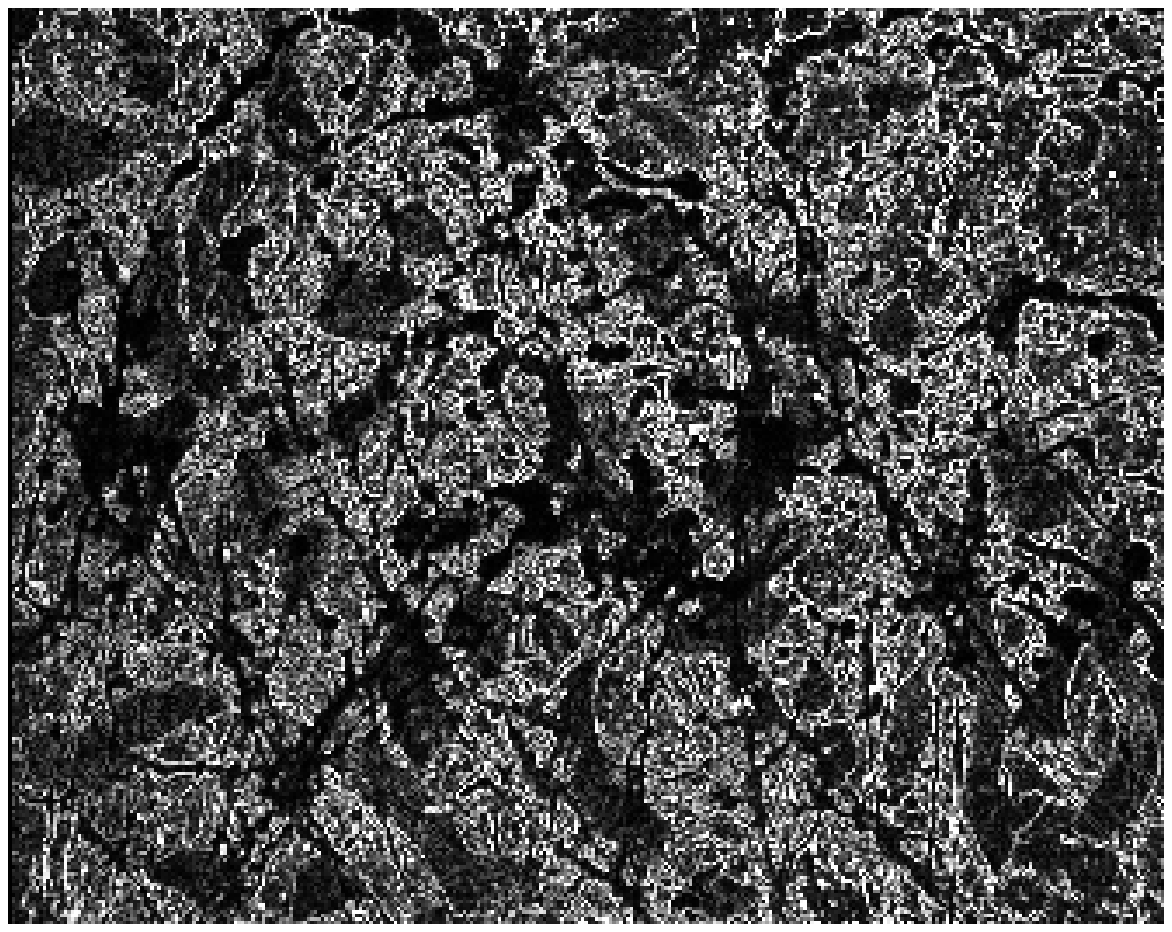}
\end{center}
\caption{
{\it Tumulte (1973)}, Les Automatistes.  (A) blob 
and (B) edge structure.
}
\label{figure2}
\end{figure}

\pagebreak
 
\begin{figure}[h] \begin{center} \leavevmode
\includegraphics[width=0.5\textwidth]{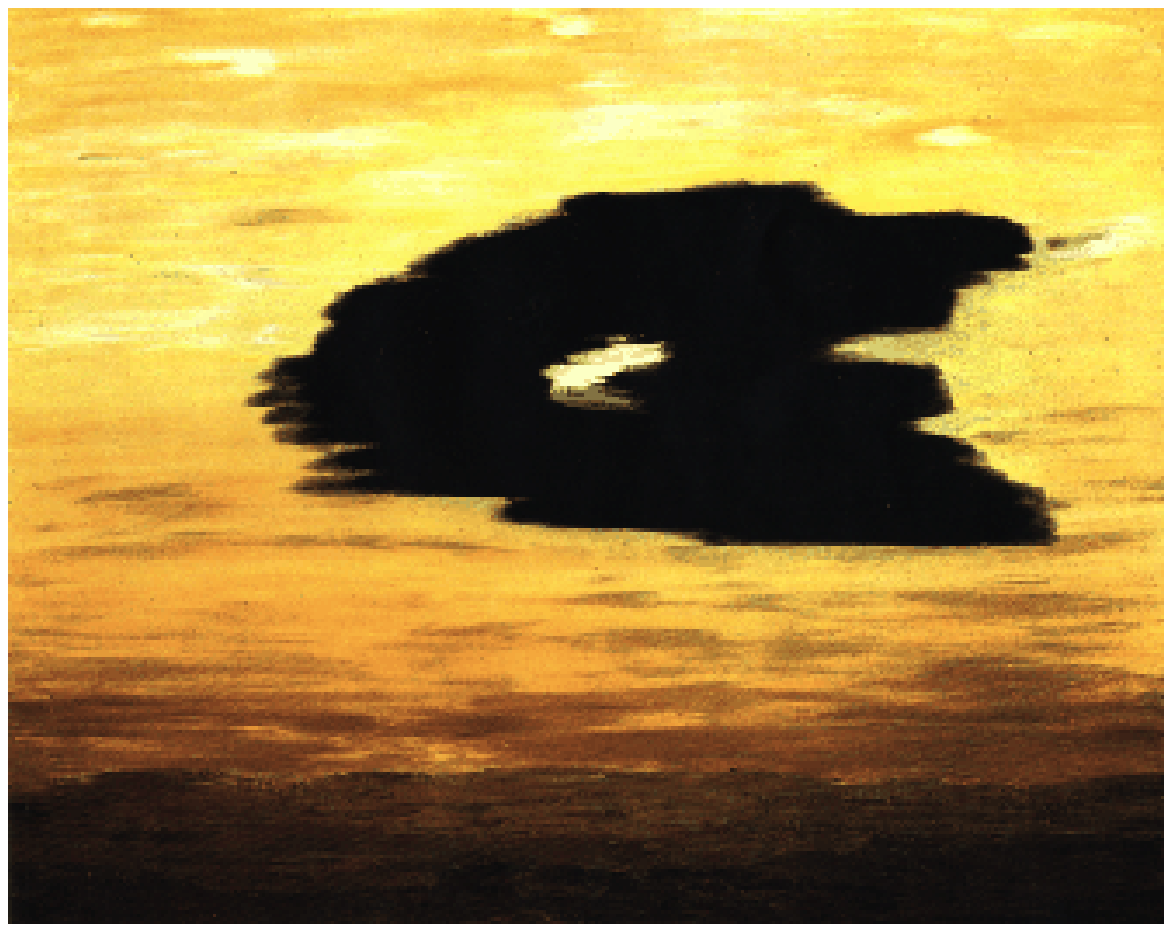}
\vskip .3cm
\includegraphics[width=0.5\textwidth]{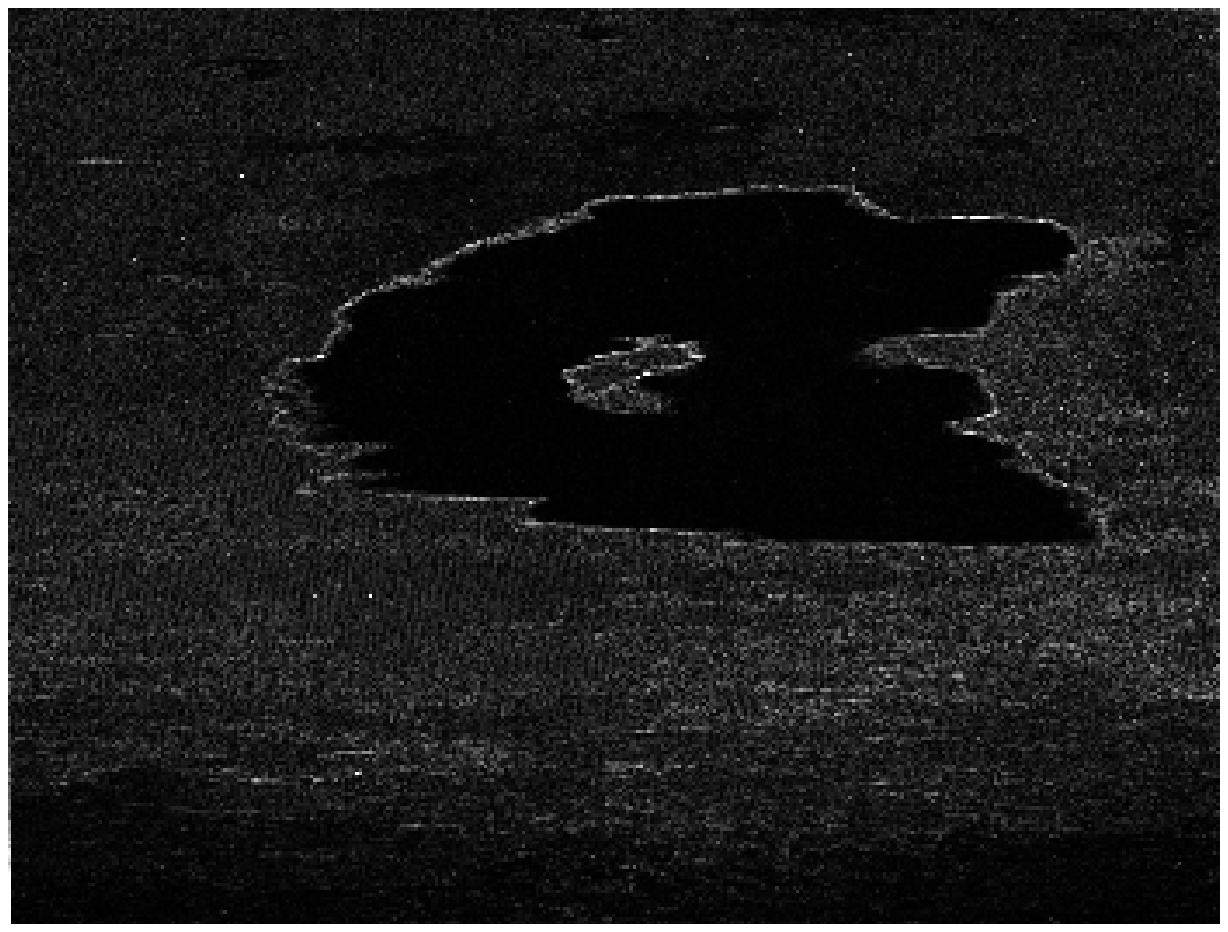}
\end{center}
\caption{
Example of Systematic Art, by Tsion Avital  (A) Raw 
image  (B) Edge structure.
}
\label{figure3}
\end{figure}

\pagebreak
 
\begin{figure}[h] \begin{center} \leavevmode
\includegraphics[width=0.7\textwidth]{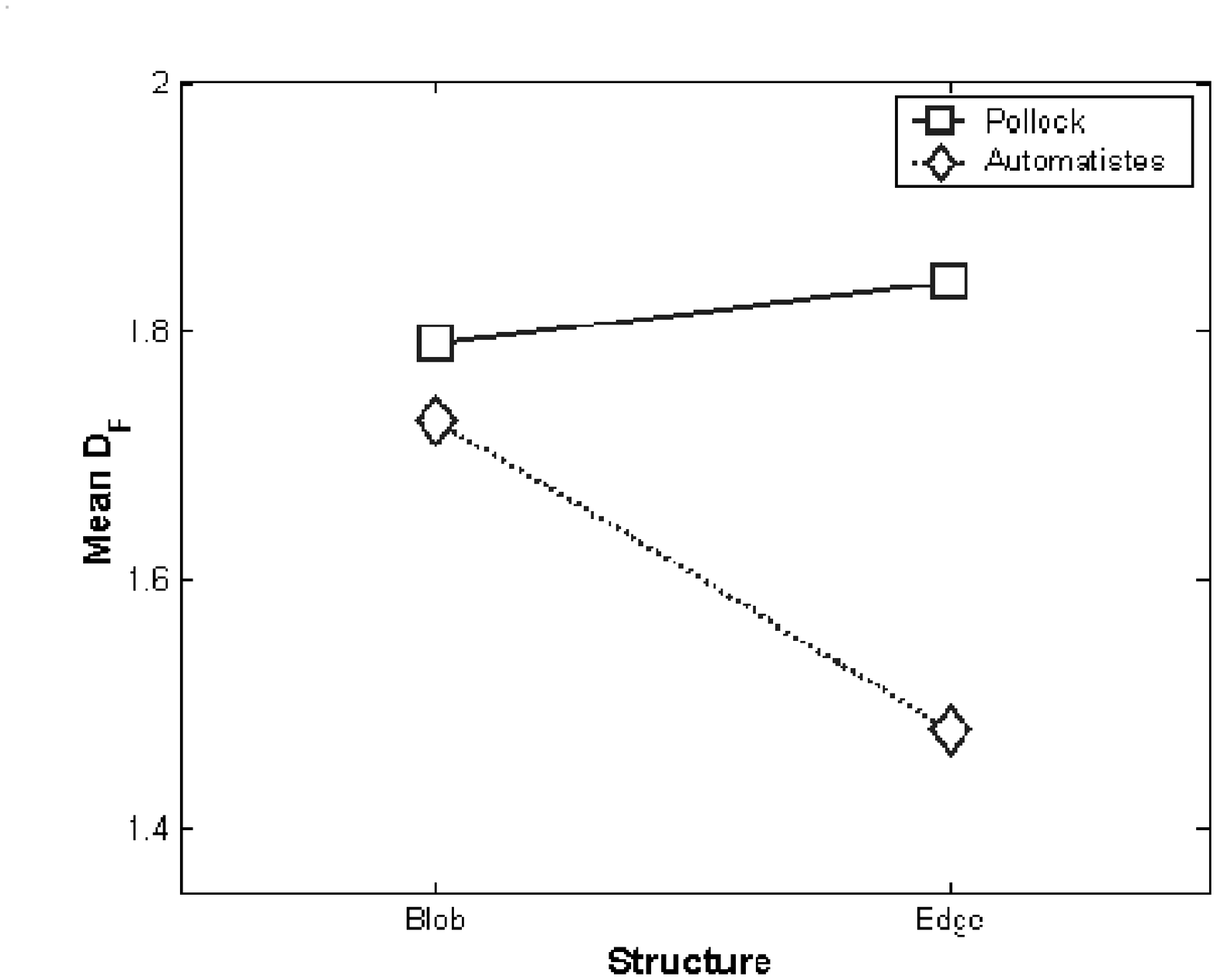}
\end{center}
\caption{
Interaction of artist and structure for average 
fractal dimension DF of blob and edge structure.
}
\label{figure4}
\end{figure}

\end{document}